\documentclass[12pt]{article}
\usepackage{graphicx,amsmath,amssymb,cite}
\usepackage{epsfig,epsf}
\usepackage{graphicx}
\usepackage{epstopdf}
\newcommand{\beq}{\begin{equation}}
\newcommand{\eeq}{\end{equation}}

\numberwithin{equation}{section}

\begin{document}

\begin{flushright}
DESY 10-061\\
SHEP-10-04\\
CERN-PH-TH/2010-091\\
\end{flushright}

\vspace*{0.5 cm}

\begin{center}

{\Large{\bf Using HERA Data to Determine the Infrared \\[1ex] Behaviour of the BFKL Amplitude }}

\vspace*{1 cm}

{\large H. Kowalski~$^1$, L.N. Lipatov~$^{2,3}$, D.A. Ross~$^4$, and G. Watt~$^5$} \\ [0.5cm]
{\it $^1$ Deutsches Elektronen-Synchrotron DESY, D-22607 Hamburg, Germany}\\[0.1cm]
{\it $^2$ Petersburg Nuclear Physics Institute, Gatchina 188300, St. Petersburg, Russia}\\[0.1cm]
{\it $^3$ II.~Institut f\"ur Theoretische Physik, Universit\"at Hamburg, D-22761 Hamburg, Germany}\\[0.1cm]
{\it $^4$ School of Physics and Astronomy, University of Southampton,\\Highfield, Southampton SO17 1BJ, UK}\\[0.1cm]
{\it $^5$ Theory Group, Physics Department, CERN, CH-1211 Geneva 23, Switzerland}
 \end{center}

\vspace*{3 cm}

\begin{center}
{\bf Abstract}
\end{center}
We determine the infrared behaviour of the BFKL forward amplitude for gluon--gluon scattering.  Our approach, based on the discrete pomeron solution, leads to an excellent description of the new combined inclusive HERA data at low values of $x$ ($<0.01$) and at the same time determines the unintegrated gluon density inside the proton, for squared transverse momenta of the gluon less than 100~GeV$^2$.  The phases of this amplitude are sensitive to the non-perturbative gluonic dynamics and could be sensitive to the presence of Beyond-the-Standard-Model particles at very high energies.

\vspace*{3 cm}

\begin{flushleft}
  November 2010 \\
\end{flushleft}

\newpage

\section{Introduction}

One of the major results from HERA~(see~\cite{H1ZEUS}, and references therein) is that the inclusive cross-section for the scattering of virtual photons against protons at low $x$ (i.e.~high energy), is dominated by the gluon density inside the proton.  This allows one to study the behaviour of the gluon density as a function of gluon momenta, i.e.~the fraction of the proton's longitudinal momentum $x$ and the transverse momentum $\mathbf{k}$.  The study of the dynamics of the gluon density is usually motivated by its importance to other physics reactions, like dijet or Higgs production at the LHC.  In addition to this merely ``utilitarian'' aspect the dynamics are very interesting because the gluon density is a fundamental quantity, comparable to black-body radiation in QED, and because gluon--gluon interactions are the source of the forces which keep matter together.

The dynamics of the gluon distribution at sufficiently low $x$ is best determined by the amplitude for the scattering of a gluon on a gluon, described  by the BFKL analysis. In this analysis the pomeron is considered as a composite state of two so-called reggeized gluons~\cite{BFKL}.  One of the salient features of the purely-perturbative BFKL analysis is the prediction of a cut-singularity with a branch point $\lambda$ leading to a low-$x$ behaviour for the gluon density of the form
\beq xg(x) \ \sim x^{-\lambda}, \eeq
with only logarithmic corrections in $x$. In leading order, $\lambda$ is given by
\beq \lambda \ = \ \frac{12\ln 2}{\pi} \alpha_s. \label{lambda} \eeq
The branch point $\lambda$ only depends on $Q^2$ through the running coupling $\alpha_s$. Experimentally, this branch point is given by the rate of rise of $F_2$ with diminishing $x$, $F_2 \sim x^{-\lambda}$.  Thus, for many years, it was claimed that BFKL analysis was not applicable to HERA data, firstly because the value of $\lambda$ obtained from (\ref{lambda}) was much larger than the observed value, and secondly because HERA  found substantial  variation of $\lambda$ with $Q^2$.

The first of these difficulties was ameliorated by the NLO contribution to $\lambda$~\cite{FL}, once the very large corrections were resummed using the collinear resummation technique~\cite{salam}.  On the other hand, the second difficulty, namely the question of the large $Q^2$ dependence of the parameter $\lambda$ remains problematic. In this paper, we will show that using a modification of the  BFKL formalism, which leads to discrete solutions (i.e.~Regge poles rather than a cut), the $Q^2$ dependence of $\lambda$ can indeed be reproduced from the BFKL amplitude. At the same time an excellent description of the low-$x$ HERA data is obtained, thereby determining the infrared behaviour of the BFKL amplitude.

\section{BFKL analysis}

The fundamental ingredient of the BFKL analysis is the amplitude for the scattering of a gluon with transverse momentum $\mathbf{k}$ off another gluon with transverse momentum $\mathbf{k^\prime}$ at centre-of-mass energy $\sqrt{s}$ which is much larger than the momentum transfer and much larger than the magnitudes of the gluon transverse momenta. In the forward case (zero momentum transfer) this amplitude, ${\cal A}(s, \mathbf{k}, \mathbf{k^\prime})$, is found to obey an evolution equation in $s$ given by
\beq \frac{\partial}{\partial \ln s} {\cal A}(s, \mathbf{k}, \mathbf{k^\prime})
 \ = \ \delta(k^2-k^{\prime \, 2})\,\delta\left(\ln\frac{s}{k k^\prime}\right) + 
 \int dq^2 {\cal K}(\mathbf{k}, \mathbf{q}) {\cal A}(s, \mathbf{q}, \mathbf{k^\prime}), 
\label{bfkl1}
\eeq
where $\cal{K}(\mathbf{k}, \mathbf{k^\prime})$ is the BFKL kernel, currently calculated to order $\alpha_s^2$~\cite{FL}. The kernel is obtained by summing all graphs which contribute to this process, but keeping only leading (and sub-leading) terms in $\ln s$. Such graphs can be drawn in terms of an effective ``gluon ladder''. The Green function evolution equation (\ref{bfkl1}) can be solved in terms of the eigenfunctions of the kernel
\beq
 \int dk^{\prime \, 2} {\cal K}(\mathbf{k},\mathbf{k^\prime}) f_\omega (\mathbf{k^\prime}) \ = \ \omega 
 f_\omega (\mathbf{k})  \label{ev2}. \eeq
In leading order and with fixed strong coupling $\alpha_s$ the eigenfunctions are parameterized by a ``frequency'' $\nu$ and are of the form
 \beq f_\omega(\mathbf{k}) \ = \  \left(k^2\right)^{i\nu-1/2}, \eeq 
with an eigenvalue, $\omega$, given by
\beq \omega \ = \ \alpha_s \chi_0(\nu), \eeq
where $\chi_0(\nu)$ is the leading order characteristic function.  The maximum value of $\omega$, at $\nu=0$, is equal to the branch point, $\lambda$.

In a recent paper~\cite{EKR}, we reported on an attempt to fit HERA data for structure functions at low values of Bjorken-$x$, using the first few discrete solutions of a modified BFKL~\cite{BFKL} equation.  The modification was proposed by one of us~\cite{lipatov86}\footnote{Discrete solutions in order to obtain discrete poles of the BFKL equation were first discussed in~\cite{BFKL,GLR,levin98}.} in order to obtain discrete poles (as opposed to the above-mentioned cut) thereby attempting to explain how QCD could, at least in principle, reproduce the very successful results of Regge theory in hadronic processes, and was the progenitor of modern string theory.

In the modified BFKL approach  the strong coupling constant $\alpha_s$ is running as  one moves away from the top or bottom of the gluon ladder. This allows the transverse momenta  of the gluons {\bf k}, which dominate the amplitude, to have a large range as one moves away from the ends of the ladder\footnote{Here the transverse momenta are controlled by the convolutions with impact factors that determine the precise process under consideration.} and results in a solution to the eigenvalue equation, (\ref{ev2}), in which  the frequencies of oscillation $\nu$  are themselves $\mathbf{k}$ dependent. In the semi-classical approximation, valid in the region
\beq \left| \frac{d}{d\ln k} \ln\left(\nu(\mathbf{k})\right) \right|   \ \ll \ \nu(\mathbf{k}), 
\label{semi1} \eeq
the solution for a given eigenvalue $\omega$ has  the form
\beq f_\omega (\mathbf{k}) \ = \ \frac{C(\nu(\mathbf{k}))}{k}
 \exp\left(2i\int^{\ln(k)} \nu(\mathbf{k^\prime}) d\ln k^\prime \right).    \label{eq7} \eeq
Here the function $\nu(\mathbf{k})$ is determined from the perturbative expansion of the kernel ${\cal K}$,
\beq \alpha_s(k^2) \chi_0(\nu(\mathbf{k}))+\alpha_s^2(k^2) \chi_1(\nu(\mathbf{k})) \ = \ \omega, 
\label{ev}
\eeq
where $\chi_0$ and $\chi_1$ are the LO and NLO characteristic functions, respectively.  We note here that in our numerical analysis, we modify $\chi_1$ following the method of Salam~\cite{salam}, in which the collinear contributions are resummed, leaving a remnant which is accessible to a perturbative analysis.  Since the RHS of eq.~(\ref{ev}) is constant it follows that the frequency $\nu$ which is the argument of these characteristic functions has to depend on $k$ in order to compensate for the $k$ dependence of the running coupling.  Eq.~(\ref{ev}) provides a match to the DGLAP equation~\cite{DGLAP} which, in the limit of small $\omega$, has an anomalous dimension $\gamma_\omega(\mathbf{k})$ related to $\nu(\mathbf{k})$ by\footnote{Note that the last term on the RHS of eq.~(\ref{anom}) arises from a conversion from an $s$-dependence determined in the BFKL approach to Bjorken-$x$, used in the DGLAP approach.}
\beq \gamma_\omega \ = \  \frac{1}{2}+i\nu(\mathbf{k}) + \frac{\omega}{2} . \label{anom} \eeq

 For sufficiently large $\mathbf{k}$, and for positive values of $\omega$, eq.~(\ref{ev}) no longer has a real solution for $\nu$ and the eigenfunctions become exponentially decreasing as opposed to oscillatory.  The transition from the real to imaginary values of $\nu(\mathbf{k})$ singles out a special value of $k = k_{\rm crit}(\omega)$, such that $\nu_\omega (k_{\rm crit})=0$.  The solutions below and above this critical momentum $k_{\rm crit}$, must be carefully matched.  Indeed, at $k=k_{\rm crit}$, eq.~(\ref{eq7}) is invalid because the coefficient $C(\nu(\mathbf{k}))$ diverges and an Airy function is used to interpolate smoothly between the two regions $k \, \ll \, k_{\rm crit}$ and $ k \, \gg \, k_{\rm crit}$~\cite{lipatov86,EKR}.  The argument of the Airy function is a function of the generalized phase,
\beq
\phi_\omega(k)= 2\int^{k_{\rm crit}}_{k} \frac{d\,k'}{k'} |\nu_\omega(k')|.   \label{phase}
\eeq 
The eigenfunction is given by 
\beq
k\, f_\omega(k) = \bar{f_\omega}(k)= \mathrm{Ai}\left(-(\frac{3}{2}\phi_\omega(k))^\frac{2}{3}\right).
\label{phase2}
\eeq
This matching of the solution to the BFKL equation in the regions $k < k_{\rm crit}$ and $k > k_{\rm crit}$ means that the BFKL equation contains more information than the DGLAP equation near $\omega=0$. Indeed, although the BFKL equation is, in principle,  an integral equation, whose solution must therefore be sampled over the entire range of $k^\prime$, the quasi-local nature of the kernel ${\cal K}(k,k^\prime)$, i.e.~the fact that the kernel only has non-negligible support where $k$ and $k^\prime$ are of the same order of magnitude, means that the kernel can be written in the form
\beq
{\cal K}(\mathbf{k},\mathbf{k^\prime}) \ = \ \frac{1}{kk^\prime} \sum_{n=0}^\infty\, c_n
 \delta^{(n)}\left(\ln(\mathbf{k}^2/\mathbf{k}^{\prime \, 2}) \right), \eeq
where the coefficients $c_n$ are given by
\beq
c_n =  \int_0^\infty  dk^{\prime \, 2} {\cal K}(\mathbf{k},\mathbf{k^\prime}) \frac{k}{k^\prime} \frac{1}{n!}
\left(  \ln(\mathbf{k}^2/\mathbf{k}^{\prime \, 2}) \right)^n   \label{cn-coeff} \eeq
rapidly falling at large $n$.  The coefficients $c_n$ are, of course, functions of $\alpha_s(k^2)$, with a perturbative expansion in $\alpha_s(k^2)$.  This leads to the BFKL equation in the pseudo-differential form
\beq
k \int dk^{\prime \, 2} {\cal K}(\mathbf{k},\mathbf{k^\prime}) f_\omega (\mathbf{k^\prime}) \ = \ 
\sum_{n=0}^\infty c_n \left( \frac{d}{d\ln(\mathbf{k}^2)}\right)^n  \bar{f_\omega} (\mathbf{k})  \ = \
\omega 
\bar{f_\omega} (\mathbf{k})  \label{bfkl-dif}. \eeq 
This means that the integral BFKL equation is equivalent to a quasi-local equation which can be cast in the form of the pseudo-differential equation
\beq 
k \int dk^{\prime \, 2} {\cal K}(\mathbf{k},\mathbf{k^\prime}) f_\omega (\mathbf{k^\prime}) \ = \ 
\chi\left( -i \frac{d}{d\ln{k^2}}, \alpha_s(k^2) \right) \bar{f_\omega}(k) \ = \ \omega
\bar{f_\omega}(k) \label{bfkl-lin}. \eeq 
In the semi-classical approximation for which $k$-dependence on $f_\omega(k)$ is such that\footnote{This condition is, in fact, equivalent to the condition of eq.~(\ref{semi1}).}
\beq \left( \frac{d}{d\ln(k)} \right)^r \bar{f_\omega}(k) \ \approx \ 
\bar{ f_\omega}(k) \left(\frac{ d \ln \bar{f_\omega}(k)}{d\ln k} \right)^r ,\eeq 
eq.~(\ref{bfkl-lin}) looks like the non-linear differential equation
\beq \chi\left( -i \frac{d \ln\bar{f_\omega}(k)}{d\ln{k^2}}, \alpha_s(k^2) \right)  \ = \ \omega 
\label{nonlinear}. \eeq
This is equivalent to the DGLAP equation written in the usual linear form
\beq \frac{d  \bar{f_\omega}(k)}{d\ln(k^2)} \ = \ 
i\nu_\omega(\alpha_s(k^2)) \bar{f_\omega}(k), \label{dglap} \eeq
where $\nu_\omega(\alpha_s(k^2))$  is the solution to
\beq  \chi\left(\nu_\omega(\alpha_s(k^2)) , \alpha_s(k^2) \right) \ = \ \omega. \label{chi-l} \eeq

Near $k=k_{\rm crit}$, where $\nu$ is small, only the second derivative term in eq.~(\ref{bfkl-dif}) is important, and the differential equation then has the form of a Schr\"odinger equation. Moreover, since in that region the difference between $\alpha_s(k^2)$ and $\alpha_s(k_{\rm crit}^2)$ is approximately linear in $\ln(k^2/k_{\rm crit}^2)$, the ``potential'' of this Schr\"odinger equation is linear and the solution is an Airy function.  Away from this region the Airy function is oscillatory for  $k \ll k_{\rm crit}$ and exponential for $k \gg k_{\rm crit}$.  The oscillatory behaviour of the Airy functions is the same as of the linear combinations of the solutions of the DGLAP equation (\ref{dglap}),
\beq
\bar{f_\omega}(k)=\exp(\pm i\phi_\omega (k)) \label{osci},
\eeq
where $i\phi_\omega (k)$ is the generalized phase of eq.~(\ref{phase}).  This means that the solution of eq.~(\ref{phase2}) provides a single interpolating function which  reproduces, to a very good approximation, the eigenfunctions of the kernel for all values of transverse momentum, $k$, with oscillation frequency given by eq.~(\ref{chi-l}).

The DGLAP equation is a linear equation which has two oscillatory solutions in the region $k \, \ll \, k_{\rm crit}$.  The BFKL equation, eq.~(\ref{bfkl-lin}), can be considered as a quantized version of the DGLAP equation (\ref{nonlinear}).  The matching of the BFKL solution in the region $k \, \sim \, k_{\rm crit}$ (which is {\it outside} the region where the semi-classical approximation is valid since the logarithm of the oscillation frequency, $\ln(\nu)$,  becomes infinite at $k_{\rm crit}$) imposes unique coefficients for the two solutions, eq.~(\ref{osci}), which cannot be obtained from the DGLAP equation alone.

Moreover, the quasi-local property of the BFKL kernel, which allows it to be recast into a  differential equation (albeit of infinite order), permits the continuation of the matched solution from large transverse momenta $ k \, \le \, k_{\rm crit}$ down to a value $ k \, \gtrapprox \Lambda_{\rm QCD}$ at which point the perturbative analysis becomes invalid.  Nevertheless, the facility to continue the solution near the infrared region means that the unknown effects from the non-perturbative sector of QCD can be encoded into information about the phase of the oscillations at some small transverse momentum.  Thus, for sufficiently low transverse momentum, $k \, \ll k_{\rm crit}$, this Airy function solution has the asymptotic behaviour
\beq
k\, f_\omega(k)  \sim \sin \left(\phi_\omega(k)+\frac{\pi}{4}\right).
\eeq
For small values of $k$ the non-perturbative effects become important and these determine the phase of the oscillations at the point $k=k_0\,  \gtrapprox \Lambda_{\rm QCD}$, denoted by $\eta$.  In Ref.~\cite{lipatov86} it was assumed that this phase is common to all eigenfunctions.  This gives rise to two boundaries,  one at $k=k_{\rm crit}$ and the other at $k=k_0$, and leads, together with the perturbative result of eq.~(\ref{phase}), to the quantization condition
\beq
\phi_\omega(k_0)=\left(n-\frac{1}{4}\right)\pi + \eta\, \pi.
\label{boundary}
\eeq

\begin{figure}[h]
\centerline{\epsfig{file=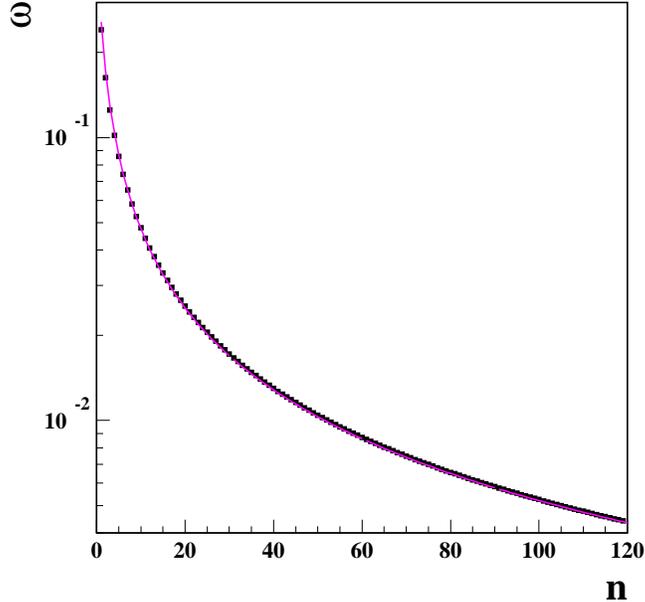, width=10 cm}}
\caption{The eigenvalues $\omega_n$ for the eigenfunctions $f_{\omega_{n}}$, with $n=1,2,\ldots,120$.  A \emph{thin line} indicates the approximate relation, $\omega_n = 0.5/(1+0.95\, n)$.   
} \label{fig-om}
\end{figure}

Consistency with both boundary conditions can only be achieved for a discrete set of eigenvalues $\omega$ and leads to a discrete set of solutions.  Figure~\ref{fig-om} shows as an example the values of $\omega_n$ for the eigenfunctions $f_{\omega_{n}}$, with $n=1,2,\ldots,120$ determined assuming that $\eta=0$, which were found by solving the equation (\ref{chi-l}) numerically.  The eigenvalues $\omega_n$ are approximately related to the eigenfunction number $n$ by a simple function
\beq \omega_n \approx \frac{0.5}{1+0.95\, n} \label{omega} .\eeq

\begin{figure}[h]
\centerline{\epsfig{file=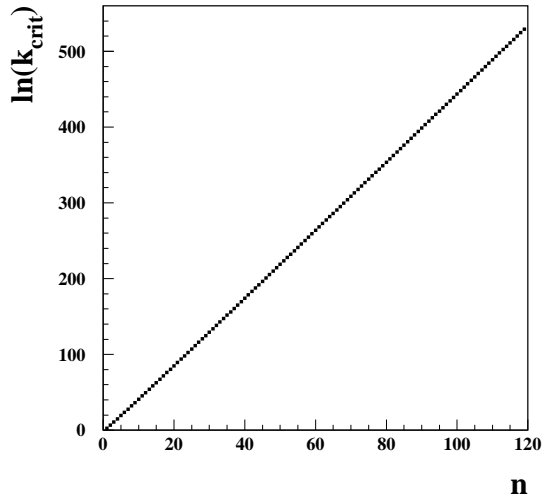, width=8 cm}}
\caption{Logarithms of the critical momenta  $k_{\rm crit}$ for the eigenfunctions $f_{\omega_{n}}$, with $n=1,2,\ldots,120$.
} \label{fig-kcrit} 
\end{figure}

\begin{figure}[h]
\centerline{\epsfig{file=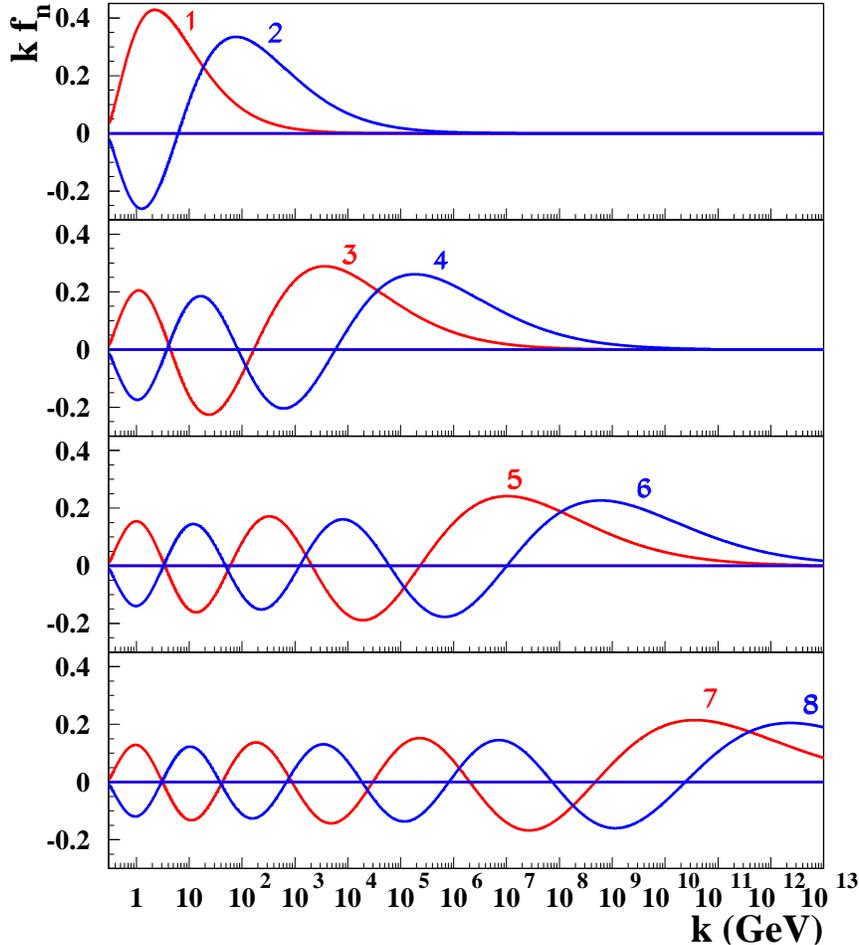, width=12 cm}}
\caption{The first eight eigenfunctions $f_{\omega_{n}}, \,\, n=1,\ldots,8$ determined at $\eta =0$.
} \label{fig-f8examp} 
\end{figure}

Figure~\ref{fig-kcrit} shows the values of the logarithms of critical momenta for the same eigenfunctions.  Note that the critical momenta are growing with increasing $n$ very fast, because the eigenfunctions oscillate, approximately, as a function of $\log(k)$.  Figure~\ref{fig-f8examp} shows the first eight eigenfunctions, $f_{\omega_{n}}, \,\, n=1\ldots8 $ determined at $\eta =0$. 

The quasi-local property of the BFKL kernel serves a further purpose.  The scattering amplitude for two gluons is obtained from the eigenfunctions  of the BFKL kernel in the multi-Regge regime, i.e.~$\ln s \, \gg \, \ln k_i$, where $\mathbf{k}_i$ is the transverse momentum of {\it any} of the gluons exchanged in the $t$-channel.  The transverse momenta of the external gluons are limited by the impact factors of the proton and photon, so that the contribution to structure functions from gluons with transverse momentum much larger than the photon virtuality, $Q$, is highly suppressed.  In the central-rapidity  region, it is possible that gluons can diffuse into gluons with much larger transverse momenta, but in this case the quasi-local property of the kernel leads to a suppression of the contribution from such gluons. Thus we are always forced to remain within the multi-Regge kinematic regime. On the other hand, since the BFKL amplitude is valid for any energy, at sufficiently large energies, since the values of $k_{\rm crit}$ are very large, there would be non-negligible contributions from gluons with large transverse momenta, above the threshold of any possible physics Beyond the Standard Model (BSM).


In~\cite{EKR}, we expressed the low-$x$ structure function of the proton, $F_2(x,Q^2)$, in terms of the discrete BFKL eigenfunctions by 
\begin{equation}
F_2(x,Q^2) \ = \ \int_x^1 dz \int \frac{dk}{k} \Phi_{\mathrm{DIS}}(z,Q,k)
  x g\left(\frac{x}{z},k\right),
 \label{f2-expr}
\end{equation}
where $x g\left(\frac{x}{z},k\right)$ denotes the unintegrated gluon density
\begin{equation}
x g(x,k) \ = \ \sum_n  \int \frac{dk^\prime}{k^\prime} \Phi_p(k^\prime)
 \left(\frac{k^\prime \, x}{k}\right)^{-\omega_n} k^2  f^*_{\omega_n}(k^\prime) f_{\omega_n}(k)
\label{un-glu}
\end{equation}
and $\Phi_p(k)$ denotes the impact factor that describes how the proton couples to the BFKL amplitudes at zero momentum transfer. The impact factor, $\Phi_{\mathrm{DIS}}(z,Q,k)$,  which describes the coupling of the virtual photon to the eigenfunctions is given in~\cite{KMS}; the dependence on $z$ reflects the fact that beyond the leading-logarithm approximation, the longitudinal momentum fraction, $x$ of the gluon differs from the Bjorken-value, determined by $Q^2$.   $\Phi_{\mathrm{DIS}}(z,Q,k)$ of Ref.~\cite{KMS} is determined taking into account kinematical constraints allowing for non-zero quark masses.

We were able to demonstrate that a good fit to data could be obtained using only the first four discrete eigenfunctions.  In that paper~\cite{EKR}, however, no attention was paid to the exact form of the proton impact factor.  In the fit, the unintegrated gluon density was expressed as 
\begin{equation}
 x g(x,k) \ = \  \sum_n  a_n x^{-\omega_n} k^{(2+\omega_n)} f_{\omega_n}(k).
 \label{xgx}\end{equation}
and only the four coefficients $a_n$ were fitted to the data.  The proton impact factor resulting from the coefficients $a_n$ turns out to be highly unstable and becomes negative for sufficiently large values of transverse momentum. 

In this paper, therefore, we fix  the proton impact factor $\Phi_p(\mathbf{k})$ in the form
\beq \Phi_p(\mathbf{k}) \ = \ A \, k^2 e^{-b k^2} \label{impact}, \eeq
which vanishes at small $k^2$ as a consequence of colour transparency, is everywhere positive and decreases for large transverse momentum, with a maximum to be determined by the fit. The particular functional form of the impact factor is not very important as long as it is positive and concentrated at the values of $k < {\cal O}(1)$ GeV. The coefficients $a_n$ are now determined from the impact factor through
\begin{equation}
a_n \ = \ \sum_n  \int \frac{dk^\prime}{k^\prime} \Phi_p(k^\prime)
 (k^\prime)^{-\omega_n}   f^*_{\omega_n}(k^\prime).
\label{a-overln}
\end{equation}

In order to do this, it turns out that it is necessary to take many more eigenfunctions because the values of the overlap integrals in eq.~(\ref{a-overln}) are diminishing  very slowly, $a_n \sim 1/\sqrt{n}$. This is due to the fact that the amplitude of the eigenfunctions is dropping like $\sim 1/\sqrt{n}$ and that all the eigenfunctions have a similar, sinusoidal, shape near $k_0$. On the other hand, the $x^{-\omega_{n}}$ enhancement of the leading contributions, eq.~(\ref{un-glu}), is weak because the values of $\omega$ are not large, even for the leading eigenfunctions, and are dropping to $\sim 0$ for larger $n$, see Fig.~\ref{fig-om}.  We will discuss this more fully at the end of Section 8.

A reasonable representation of a realistic impact factor of the proton, in terms of the eigenfunctions, requires a modification of the assumption of a common phase at some low value of transverse momentum.  In Section~\ref{sec:data}, we demonstrate that such phases can be obtained with good accuracy by confronting the discrete eigenfunctions with the HERA data.

\section{Dependence of the infrared phase on eigenfunction number}

In this section, we explain why the ansatz of a common infrared phase, $\eta$ at $k=k_0$ is {\it not} compatible with a ``reasonable'' form for the proton impact factor such as the one suggested in eq.~(\ref{impact}). We begin by repeating, briefly, the argument in~\cite{lipatov86} for a constant $\eta$, which was given within the context of the LO BFKL kernel only. In this approximation the eigenvalue equation (with running coupling in the semi-classical approximation) is
\beq \alpha_s(k^2) \int d k^{\prime \, 2} {\cal K}_0(\mathbf{k}, \mathbf{k}^\prime) f_\omega(\mathbf{k^\prime})
 \ = \ \omega f_\omega(\mathbf{k}). \eeq
In the infrared limit, where $\omega \ll \alpha_s(k^2)$ this may be approximated by
\beq
\int d k^{\prime \, 2} {\cal K}_0(\mathbf{k}, \mathbf{k}^\prime) f_{\omega}(\mathbf{k^\prime}) \ = \ 0 \eeq
whose solution is of the form
 \beq f_\omega(\mathbf{k}) \ = \ \frac{1}{k} \sin \left(2\nu_0 \ln(k/k_0) + \eta\pi \right) ,\label{eta} \eeq
irrespective of the value of $\omega$.  Here $\nu_0 \ \sim 0.64$, is the solution to $$ \chi_0(\nu_0) \ = \ 0, $$$\chi_0$ being the LO characteristic function.

A couple of comments are in order:
\begin{enumerate}
\item The argument does not lend itself to an analysis of the kernel which includes the NLO contribution in the infrared region where $\alpha_s(k^2)$ becomes large. 
\item For the higher eigenfunctions for which $\omega$ is indeed negligible compared with the running coupling at $k_0$, the phase $\eta$ in the solution (\ref{eta}) can depend on $\omega$ because the NNLO corrections are not taken into account.  In addition, non-perturbative effects are expected to determine the behaviour of the BFKL amplitude for $k \, < \, k_0$, but there is no {\it a priori} reason to suppose that such an infrared behaviour is given by a common phase for all the eigenfunctions---and indeed we show that this cannot be the case.
\end{enumerate}

\begin{figure}
\centerline{\epsfig{file=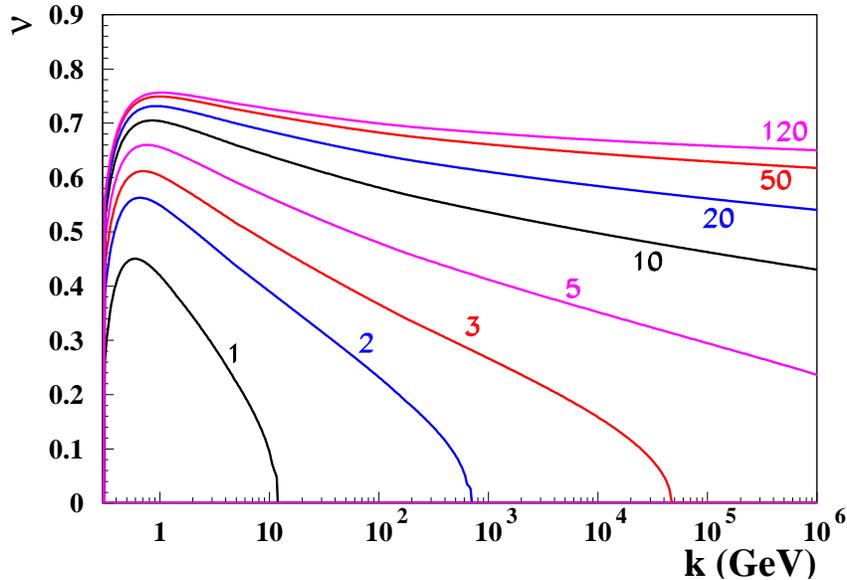, width=12cm}} 
\caption{The frequencies $\nu(k)$ for the eigenfunctions $1,2,3,5,10,20,50$ and $120$.}
\label{nuv-pap} \end{figure}

The salient feature of the eigenfunctions we obtain is the fact that above the first two eigenfunctions the value of $k_{\rm crit}$, at which the frequency of oscillation vanishes, is many orders of magnitude above where we are fitting data, see Fig.~\ref{fig-kcrit} and Fig.~\ref{fig-f8examp}.  This means that in the fit region, we are essentially fitting the proton impact factor to oscillatory functions, in $\ln(k/k_0)$, whose frequency actually varies very little as we scan through the relevant range of $k$, as can be seen in Fig.~\ref{fig-f8examp}.  For $n>2$, the difference between the eigenfunctions $n$ and $n+1$ is that the eigenfunction $n+1$ has one more oscillation than the eigenfunction $n$.  However, in the region of HERA data, $k<{\cal O}(30)$ GeV, both eigenfunctions oscillate in a similar way. This is also directly shown in Fig.~\ref{nuv-pap}, where we plot  the frequencies $\nu(k)$ for the eigenfunctions 1, 2, 3, 5, 10, 20, 50 and 120. We see that frequencies $\nu$ vary  between 0 and 0.75 only.  Thus we are effectively performing a Fourier analysis, for which a (generally infinite) range of frequencies is necessary, with a very limited frequency range.

Now, since in this Fourier transform of the impact factor, the quantity $\nu$ and $\ln(k/k_0)$ are conjugate variables, a Fourier transform with a limited range, $\Delta\nu \, \sim \, 0.75$ can only be achieved if the function has a correspondingly broad range, $\Delta\ln(k/k_0)$, in the variable $\ln(k/k_0)$ where
$$ \Delta \nu \, \Delta \ln(k/k_0) \ \geq \pi .$$
This means that the proton impact factor would have to have non-negligible support up to a value $k_{\rm max}$ of transverse momentum, where
\beq 0.75 \ln\left(\frac{k_{\rm max}}{k_0}\right) \ \sim \pi .\eeq
This gives $k_{\rm max} \, \sim \,  20$  GeV, which is much larger than one would want for a proton impact factor, which is expected to peak at a few hundred MeV and decrease rapidly thereafter.

 The BFKL kernel possesses a complete set of eigenfunctions and therefore one would expect to be able to expand any function (e.g.~any proton impact factor) in terms of these functions. However, the eigenfunctions which display oscillations with frequency greater than the limit discussed above would have {\it negative} eigenvalue, $\omega$.  Such negative eigenvalue solutions are expected to be of negligible importance at sufficiently low $x$ (they contribute {\it positive} powers of $x$).  They are also very sensitive to unknown higher order corrections to the BFKL kernel, as the perturbative expansion in a power series in $\alpha_s$ is strictly only valid in the limit $\alpha_s \, \leq \omega$.  Furthermore there is no reason to assume that for negative $\omega$ there exists a branch point in $\nu$ which leads to a discrete spectrum.  We therefore seek an expansion for the proton impact factor in terms of the (positive $\omega$) discrete subset of eigenfunctions, which can be achieved provided the boundary conditions of the proton impact factor are compatible with the physical properties of this subset of eigenfunctions.  The assumption of a constant infrared phase $\eta$ at some low $k=k_0$ imposes a further boundary condition on the set of eigenfunctions.  The consequent upper limit on the oscillation frequencies of the eigenfunctions obtained means that this further boundary condition is not compatible with a proton impact factor that decreases on the scale of a few hundred MeV.  We are therefore led inevitably to the conclusion that we must release entirely the constraint on $\eta$ and allow this phase to depend on eigenfunction number.  This would shift the relative phases of oscillations of various eigenfunctions drawn in Fig.~\ref{nuv-pap} and allow us to present our impact factor in the form of their linear combination.  We use a fit to HERA  data, to pin down the dependence of the infrared phase on the eigenfunction number and thereby understand the BFKL amplitude for gluon--gluon scattering for small values of transverse momentum, where perturbation theory is no longer valid.

\section{Non-Hermitian kernel}

The introduction of running coupling into the BFKL kernel, expanded perturbatively as
\beq {\cal K}(\mathbf{k}, \mathbf{k^\prime}) \ = \ \alpha_s  {\cal K}_0(\mathbf{k}, \mathbf{k^\prime})
 + \alpha_s^2  {\cal K}_1(\mathbf{k}, \mathbf{k^\prime}) \ + \cdots \eeq
begs the question as to which transverse momentum ($\mathbf{k}$ or $\mathbf{k^\prime}$) should be used to determine the running.  In~\cite{trimvirate} a so-called ``triumvirate'' algorithm is employed, which has the advantage of being symmetric in $\mathbf{k}$ and $\mathbf{k^\prime}$ and therefore preserved the Hermiticity of the kernel.  Another possibility to reach the Hermiticity is to use the similarity transformation~\cite{FL} 
$$K(k,k^\prime) \rightarrow U(k)\, K(k,k^\prime)\, U^{-1}(k^\prime)$$
where $U(k)$ depends on $\alpha_s(k^2)$. However, because of the additional dependence of $\alpha_s(k^2)$ in both methods the number of essential eigenfunctions grows substantially, leading to significant technical difficulties.

In this analysis it was found that it is simpler to work with the set of eigenfunctions which are not completely orthogonal
\beq \int d^2 \mathbf{k} \, f_m(\mathbf{k}) f^*_n(\mathbf{k}) \ = \ {\cal N}_{mn} \ \neq \ \delta_{mn}, \label{nhort} \eeq
where, to simplify the notation, we write in the following $f_{\omega_{n}}=f_n$.  We make the assumption that these eigenfunctions form a complete set of states (subject to a given boundary condition at $k=k_0$) so that we may write
 \beq  \delta^2(\mathbf{k} - \mathbf{k^\prime} )   \ = \ 
    \sum_{n,m} f_n(\mathbf{k}) {\cal N}^{-1}_{nm} f^*_m(\mathbf{k^\prime}). \label{complete} \eeq
Then the (forward) BFKL equation for the scattering amplitude of a gluon with transverse momentum $\mathbf{k})$ off a gluon with transverse momentum $\mathbf{k}^\prime$ and centre-of-mass energy $\sqrt{s} \  (\gg k,k^\prime)$, namely
\beq \frac{\partial}{\partial \ln s } {\cal A}(s,\mathbf{k},\mathbf{k^\prime})
 \ = \ \delta^2(\mathbf{k}-\mathbf{k^\prime})\,\delta\left(\ln\frac{s}{k k^\prime}\right) + \int d^2\mathbf{q} \,
{\cal K}(\mathbf{k}, \mathbf{q})  \,  {\cal A}(s,\mathbf{q},\mathbf{k^\prime}) \label{bfkl} \eeq
 is given by
  \beq 
  {\cal A}(s, \mathbf{k}, \mathbf{k^\prime})  \ = \ \sum_{m,n} f_m(\mathbf{k}) {\cal N}^{-1}_{mn} 
 f_n(\mathbf{k^\prime}) \left(\frac{s}{k k^\prime}\right)^{\omega_m}. 
 \label{ampl-nh}
 \eeq

\begin{figure}[h]
\centerline{\epsfig{file=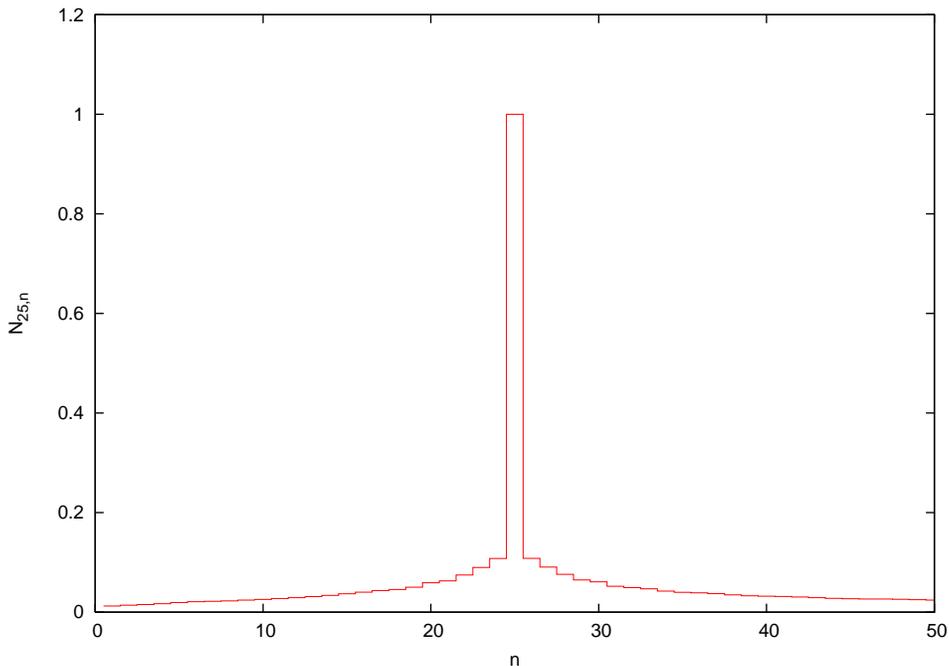, width=9 cm, angle=270}}
\caption{The overlaps ${\cal N}_{mn}$, given by eq.~(\ref{nhort}), for $m=25$.} \label{fig1} \end{figure}

As a demonstration, in Fig.~\ref{fig1} we show the overlaps ${\cal N}_{mn}$ for $m=25$.  We can see that the off-diagonal elements are much smaller than the diagonal term, as expected from the fact that the region where $k$ and $k^\prime$ are substantially different generally contributes little to the amplitude.  This is a consequence of the fact that in this region the semi-classical approximation works fairly well and therefore we can neglect the non-commutativity of $\alpha_s(k^2)$ and $d/d\ln(k^2)$.  However, there are many such overlaps and in our analysis we probe a region in which $k$ is relatively large, being controlled by the $Q^2$ of the structure functions, whereas $k^\prime$  is controlled by the proton impact factor and is expected to be relatively small. Thus these off-diagonal components of the amplitude turn out to have a significant effect on the quality of the fit obtained.

\section{Properties of the Green function}

We can define the pomeron Green function  in accordance with eq.~(\ref{ampl-nh}) as 
\beq
G_y(\mathbf{k},\mathbf{k'})=\sum_{mn}f_m(\mathbf{k})N_{mn}^{-1}f_n(\mathbf{k'})
e^{\omega _ny}\,,
\eeq
where $y$ denotes  the relative gluon rapidity.  It  satisfies the integral equation
\begin{equation}
G_y(\mathbf{k},\mathbf{k'})=\int d^2 \mathbf{k''} G_{y'}(\mathbf{k},\mathbf{k''})G_{y-y'}(\mathbf{k''},\mathbf{k'}) \label{non-lin}
\end{equation}
for an arbitrary rapidity $y'$ in the interval $0<y'<y$.  The quasi-locality and the semi-classical approximation ensures that in the integral of eq.~(\ref{non-lin}), at fixed $\mathbf{k},\mathbf{k'}$ and $y$, the essential values of $\mathbf{k''}$ are restricted from above. This is a non-trivial property  of the Green function but not of the eigenfunctions themselves, because in the orthogonality condition (\ref{nhort}) large values of $\mathbf{k}$, of the order of $k_{\rm crit}$ at large $n$ or $m$, contribute significantly.

\begin{figure}[h]
\centerline{\epsfig{file=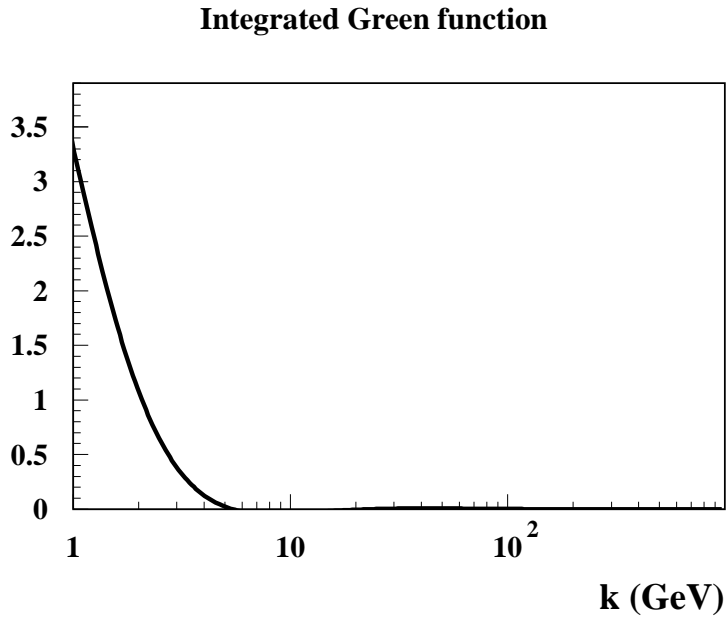, width=12 cm}}
\caption{Distribution of the momentum $k$ in the Green function,  $G_y(\mathbf{k},\mathbf{k'})$, integrated over $k'$ with the proton impact factor at $y=\ln (s/k^2)=\ln (1/x= 10^{3})$.} \label{greenf} 
\end{figure}

To verify that this quasi-locality property is also present in our numerical evaluation we show in Fig.~\ref{greenf} the distribution of the momentum $\mathbf{k}$ in the Green function, $G_y(\mathbf{k},\mathbf{k'})$, integrated over $\mathbf{k'}$ with the proton impact factor as in eq.~(\ref{intprot}).

Figure~\ref{greenf} shows that indeed, when we limit $k'<1$~GeV (by integrating with the proton impact factor) the contributions of large  momenta $k$ are strongly suppressed for $k>3$~GeV. Therefore, the diffusion into large momenta, which could seem natural for large $n$ eigenfunctions, is strongly suppressed.   On the other hand, we will show in subsequent sections that large $n$ eigenfunctions make significant contributions and are essential for the description of data with high quality.

\section{Comparison with HERA data} \label{sec:data}

In this section we determine the $\eta$--$n$ relation and  the free parameters of the proton impact factor by fitting  the new, combined, HERA data for the structure function  $F_2$~\cite{H1ZEUS}.  In analogy to  eqs.~(\ref{f2-expr}) and (\ref{un-glu}) we obtain the expression for $F_2$ by performing convolutions of the non-Hermitian BFKL amplitude, eq.~(\ref{ampl-nh}), with the photon and proton impact factors:

\beq {\cal A}^{(U)}_n \ \equiv \
 \int_x^1 \frac{d\xi}{\xi} \int \frac{dk}{k} \Phi_ {\mathrm {DIS}}(Q^2,k,\xi) \left(\frac{ \xi k}{x}\right)^{\omega_n}
   f_n(\mathbf{k}), \eeq

\beq {\cal A}^{(D)}_m \ \equiv \
  \int \frac{dk^{\prime}}{k^{\prime}} \Phi_p(k^\prime) \left(\frac{ 1}{k^\prime}\right)^{\omega_m}
   f_m(\mathbf{k}^\prime). \label{intprot}\eeq

The expression for the structure function is then
\beq F_2(x,Q^2) \ =  \ 
 \sum_{m,n}  {\cal A}^{(U)}_n  {\cal N}^{-1}_{nm}  {\cal A}^{(D)}_m. \eeq
This expression can be compared directly to data after the $\eta$--$n$ relation is assumed. In the search for this relation we were guided by the principle of simplicity and some analogy to the Balmer series.  In the Balmer series the energies of the quantum levels have a simple dependence on the principal quantum numbers $n$. In the QCD version of the Regge theory developed here the BFKL  equation is considered to be analogous to the Schr\"odinger equation for the wavefunction of the pomeron.  The BFKL kernel corresponds to the Hamiltonian and the eigenvalues $\omega$  to the  energy  eigenvalues.  The BFKL equation determines the dependence on $\omega=\omega_n$ provided the boundary conditions at small $k \sim \Lambda_{\rm QCD}$ are defined by specifying the non-perturbative phase $\eta$  which could depend on $n$.  We have tried several two-parameter forms for the $n$-dependence of $\eta$,  but we found that a simple one-parameter functional form
\beq 
\eta \ =  \eta_0 \left(\frac{n-1}{n_{\rm max}-1}\right)^{\kappa}
\label{eta-n}
\eeq
gives the best fits to data in terms of $\chi^2/N_{df}$, where $N_{df}\equiv N_{\rm dat}-N_{\rm par}$ with $N_{\rm dat}$ the number of fitted data points and $N_{\rm par}$ the number of fitted parameters.  Here, $n_{\rm max}$ denotes the maximum number of eigenfunctions, $\kappa$ is a free parameter and $\eta_0$ denotes the range of $\eta$ values used in the fit. The form of this equation guarantees that the phase of the first eigenfunction is zero. The value of $\eta_0$ was determined from the requirement that the value of the overlap integral ${\cal A}^{(D)}_m $ should be zero when $m=n_{\rm max}$.  This condition minimizes the contributions of the eigenfunctions with $n>n_{\rm max}$, because the frequencies $\nu(k)$ of the higher $n$ eigenfunctions are very similar in the low $k$ region.  We have investigated other functional forms for the $n$-dependence of $\eta$ which include non-zero phases for the leading eigenfunction and combinations of both negative and positive powers of $n$.  However, we find that the simple form of eq.~(\ref{eta-n}) gives the best fit.  In the fit, only three parameters were determined from data: the power $\kappa$,  the normalization constant $A$ and the suppression slope $b$ of the proton impact factor, eq.~(\ref{impact}).

The separate H1 and ZEUS inclusive cross-section measurements taken in the period 1994--2000 have recently been \emph{combined} to improve accuracy~\cite{H1ZEUS}.  We fit the 128 measured $F_2$ data points for neutral-current $e^+p$ scattering with cuts $Q^2>4$~GeV$^2$ and $x<0.01$.  Statistical and systematic uncertainties are added in quadrature.  The overall data normalization uncertainty of 0.5\% is absorbed into the fitted $A$ parameter in the proton impact factor, eq.~(\ref{impact}).  We take an input running coupling $\alpha_s(M_Z)=0.1176$ with heavy flavour thresholds at $m_c=1.40$~GeV, $m_b=4.75$~GeV and $m_t=175$~GeV.  Massive charm and bottom quark contributions are included in the photon impact factor~\cite{KMS} with the same values of $m_c$ and $m_b$.

The computation of large numbers of eigenfunctions together with their overlaps poses a considerable numerical problem.  Within the adopted computational precision we were able to compute up to 150 eigenfunctions and up to 30 overlaps of each eigenfunction with their closest neighbours and obtain a value of $F_2$ with a relative precision of $\sim1\% $.  This precision is necessary to match the precision of the measured $F_2$ values of $\sim2\% $.
\begin{table}
(a) Fits with cuts of $Q^2>4$ GeV$^2$ and $x<0.01$:\\
\begin{center}
\begin{tabular}{|c|r|r|c|c|c|} \hline
\, \,\, \,$ n_{\rm max}$  \,\,\, &\;\;\; $\chi^2/N_{df}(x<0.01)\;\;\;$  & \;\; $\chi^2/N_{\rm dat}(x<0.001) $\;\;  &\,\,\,$\kappa$\,\,\, &\,\,\, $A$ \,\,\,&\,\,\, $b$\,\,\,\\ \hline \hline
1   & 9792 /125 = 78.3 & 2123 /43 = 49.4      & \;\; --- \;\; &  \;\;  156 \;\; & \;\; 30.0 \;\;     \\ \hline
5   & 349.8 /125 = 2.80 & 88.8 /43 = 2.07   & \;\;3.78\;\; &  \;\;  $3.1\cdot 10^{6}$ \;\; & \;\; 78.0 \;\;     \\ \hline
20   & 286.5 /125 = 2.29 & 83.3 /43 = 1.94     & \;\;0.96\;\; &  \;\;  632 \;\; & \;\; 15.8 \;\;     \\ \hline
40   & 193.3 /125 = 1.55 & 54.9 /43 = 1.28    & \;\;0.84\;\; &  \;\;  2315 \;\; & \;\; 23.2 \;\;     \\ \hline
60  & 163.3 /125 = 1.31 & 44.8 /43 = 1.04      & 0.78 &   3647   & 25.6       \\ \hline
80   & 156.5 /125 = 1.25 &  43.5 /43 = 1.01    & 0.73 &  3081   & 24.4     \\  \hline
100  & 149.1 /125 = 1.19 &  41.3 /43 = 0.96   & 0.69 &  2414   & 22.8  \\  \hline
120  & 143.7 /125 = 1.15 &  39.2 /43 = 0.91  & 0.66 &  2041   & 21.8   \\  \hline
\end{tabular}
\end{center}
\vspace{1cm}
(b) Fits with cuts of $Q^2>4$ GeV$^2$ and $x<0.001$:\\
\begin{center}
\begin{tabular}{|c|r|c|c|c|} \hline
\, \,\, \,$ n_{\rm max}$  \,\,\, & \;\; $\chi^2/N_{df}(x<0.001) $\;\;  &\,\,\,$\kappa$\,\,\, &\,\,\, $A$ \,\,\,&\,\,\, $b$\,\,\,\\ \hline \hline
1   & 894 /40 = 22.4      & \;\; --- \;\; &  \;\;  193 \;\; & \;\; 29.6 \;\;     \\ \hline
5   & 55.4 /40 = 1.39   & \;\;3.65\;\; &  \;\;  $4.8\cdot 10^{6}$ \;\; & \;\; 81.5 \;\;     \\ \hline
20   & 52.6 /40 = 1.31     & \;\;1.09\;\; &  \;\;  $1.1\cdot 10^{4}$ \;\; & \;\; 33.7 \;\;     \\ \hline
40   & 35.5 /40 = 0.89    & \;\;1.00\;\; &  \;\;  $9.4\cdot 10^{5}$ \;\; & \;\; 68.2 \;\;     \\ \hline
60  & 27.6 /40 = 0.69      & 0.91 &   $1.1\cdot 10^{6}$   & 68.2       \\ \hline
80   &  27.1 /40 = 0.68    & 0.86 &  $9.3\cdot 10^{5}$   & 66.9     \\  \hline
100  &  27.3 /40 = 0.68   & 0.82 &  $7.0\cdot 10^{5}$   & 64.6  \\  \hline
120  &  27.8 /40 = 0.70  & 0.79 &  $6.0\cdot 10^{5}$   & 63.6   \\  \hline
\end{tabular}
\end{center}
\caption{The qualities of fits, $\chi^2/N_{df}$, using up to $n_{\rm max}$ eigenfunctions with a cut of $Q^2>4$~GeV$^2$ and either (a)~$x<0.01$ (128 data points) or (b)~$x<0.001$ (43 data points).  In the upper table (a) we also show the contribution to $\chi^2$ given by the 43 data points with $x<0.001$, \emph{without} refitting the parameters.  The improvement in $\chi^2$ obtained by refitting the parameters can be seen in the lower table (b).  The fits use three parameters: the dimensionless power $\kappa$ and the two parameters, $A$ and $b$, which determine the properties of the proton impact factor, and are both given in units of GeV$^{-2}$.  In the photon impact factor four flavours are taken into account.  The value of $\eta_0$ was determined to be $\eta_0=-0.9$.  The fit with $n_{\rm max}=1$ eigenfunction was performed with the optimized constant phase (a)~$\eta=-0.71$ or (b)~$\eta=-0.69$.  The fits with one, five and 20 eigenfunctions were performed with one, five and 20 overlaps, respectively.  All other fits were performed with the maximum number of available overlaps, i.e.~30.
\label{table:eval1}}
\end{table}

\begin{figure}[h]
\centerline{\epsfig{file=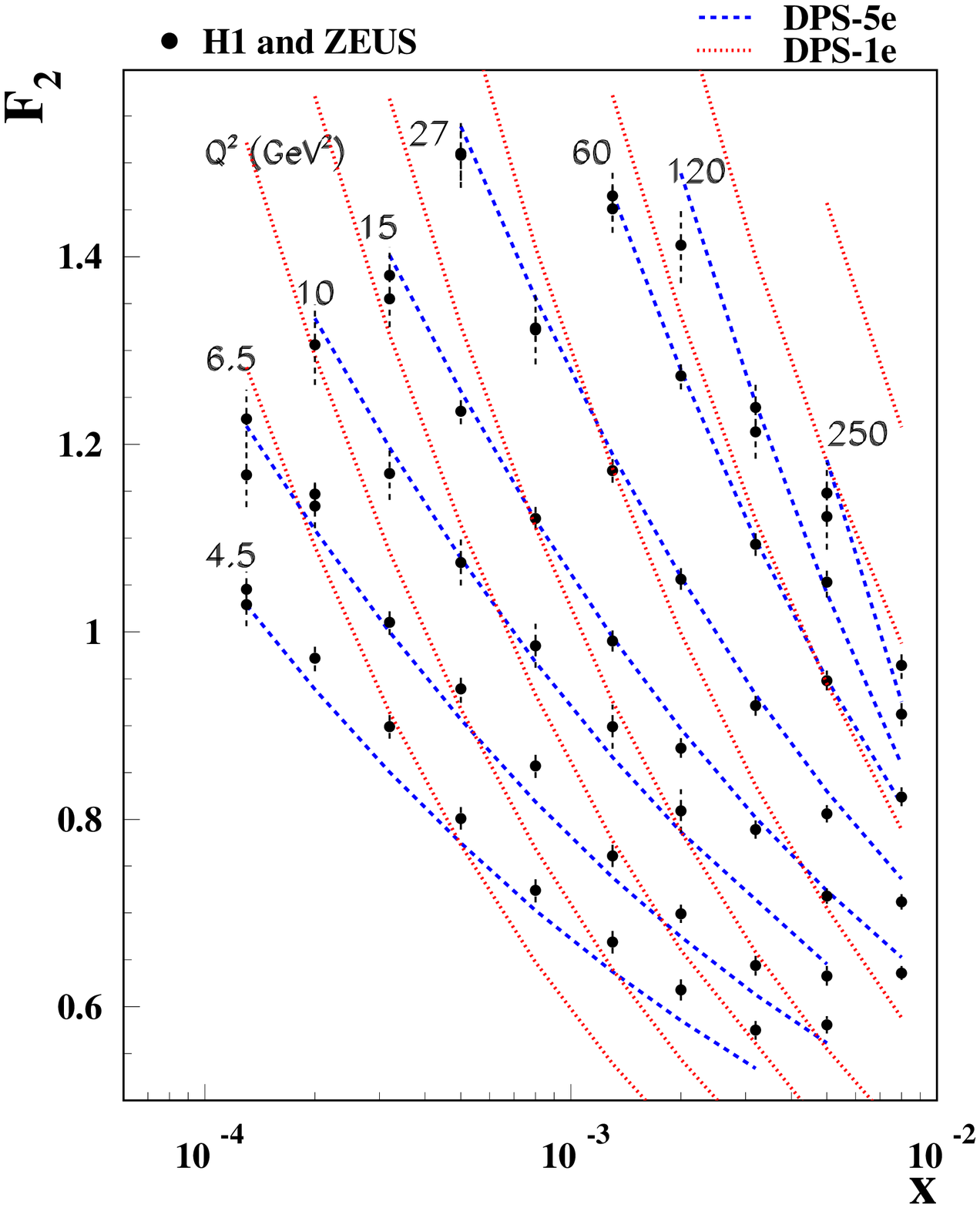, width=11 cm}}
\caption{The results of the fit, performed with one (\emph{dotted line}) and five (\emph{dashed line}) eigenfunctions, compared to a subsample of the low-$x$ HERA data~\cite{H1ZEUS}.} 
\label{globfit51} 
\end{figure}
The results of the Discrete Pomeron Solution (DPS) fit are shown in Table~\ref{table:eval1}(a) as a function of the maximal number of eigenfunctions used in the fit, $n_{\rm max}$.  The fit with only the first (leading) eigenfunction gives a very bad fit with $\chi^2/N_{df} \approx 80$. The resulting $F_2$ is even qualitatively in disagreement with data.  The fit with the first five eigenfunctions starts to reproduce qualitatively the main trends of the data although the quality of the fit is still poor, $\chi^2/N_{df} \approx 3$, see Fig.~\ref{globfit51}.  The addition of more eigenfunctions improves the fit quality substantially until, with 120 eigenfunctions, a $\chi^2/N_{df} \approx 1.2 $ is achieved.  Table~\ref{table:eval1}(a) shows an unexpected property of the discrete solution of the BFKL equation: the high quality of a fit requires a large number of eigenfunctions.  In the third column of Table~\ref{table:eval1}(a) we show the contribution to the $\chi^2$ from the subset of 43 data points with $x<0.001$, where the BFKL formalism is most applicable and the fit quality is somewhat better than in the whole region of $x<0.01$.  In Table~\ref{table:eval1}(b) we show the fit quality if \emph{only} the 43 data points with $x<0.001$ are included in the fit.  In this restricted region, convergence is faster with an increasing number of eigenfunctions, as expected, but still at least 40 eigenfunctions are needed for a good fit, then convergence is reached after 60 eigenfunctions.  The parameters determined in the fit for $x<0.001$ are similar to the parameters obtained for $x<0.01$.  However, they are less constrained when using only 43 data points (for $x<0.001$) compared to 128 data points (for $x<0.01$), therefore we choose a cut of $x<0.01$ for the following results, at the expense of requiring more eigenfunctions for an acceptable fit quality.  See also the discussion below.

In Table~\ref{table:eval2} we show the qualities of fits as a function of the number of overlap integrals, $n_{\rm overl}$, used in the fit.
\begin{table}
\begin{center}
\begin{tabular}{|c|c|c|c|c|} \hline
\, \,\, \,$ n_{\rm overl}$  \,\,\, & \;\;\;\; $\chi^2/N_{df} $\;\;\;\;  & \,\,\,$\kappa$\,\,\, &\,\,\, $A$ \,\,\,&\,\,\, $b$\,\,\,\\ \hline \hline
0   & 354.6 /125 = 2.84      & \;\;0.41\;\; &  \;\;  7.80 \;\; & \;\; 1.40 \;\;     \\ \hline
10  & 206.9 /125 = 1.66      & 0.50       &   69.1   &    5.83    \\ \hline
20   & 150.8 /125 = 1.21      & 0.60      &  444.4   & 13.5     \\  \hline
30  & 143.7 /125 = 1.15     & 0.66        &  2041   & 21.8   \\  \hline
\end{tabular}
\end{center}
\caption{The qualities of fits using up to $n_{\rm overl}$ overlap integrals, and the corresponding parameters of the fits, with $\eta_0=-0.9$ and four flavours in the photon impact factor.
The parameters $A$ and $b$ are both given in units of GeV$^{-2}$.
\label{table:eval2}}
\end{table}
All fits were performed with the same number of available eigenfunctions, $n_{\rm max}=120$.  The table shows that the quality of fits improves rapidly when the overlap integral correction is included.  The fits have only a very small sensitivity to the value of $\eta_0$, not more than 1 or 2 units in $\chi^2$, therefore $\eta_0=-0.9$ was used in all fits.

The fits shown in Table~\ref{table:eval1} and \ref{table:eval2} were made, as in our previous paper~\cite{EKR}, using four quark flavours in the photon impact factor.  Since the contribution of the bottom quark, although small, is present in HERA data we included it in our final fit. The results of the fit performed with 120 eigenfunctions, 30 overlaps and five flavours are shown in Table~\ref{table:eval3}.\footnote{We note that the value of the parameter, $b$, is such that the proton impact factor eq.~(\ref{impact}) is peaked at $ k \sim \Lambda_{\rm QCD}$, as expected.}

\begin{table}
\begin{center}
\begin{tabular}{|c|c|c|c|} \hline
 \;\;\;\; $\chi^2/N_{df} $\;\;\;\;  & \,\,\,$\kappa$\,\,\, &\,\,\, $A$ \,\,\,&\,\,\, $b$\,\,\,\\ \hline \hline
 \;\;154.7 /125 = 1.24 \;\;    & \;\;\;  0.65\;\;\;       & \;\;\; 1660 \;\;\;   & \;\;\; 20.6 \;\;\;   \\  \hline
\end{tabular}
\end{center}
\caption{The parameters of the final fit performed with 120 eigenfunctions and 30 overlaps, with $\eta_0=-0.9$ and five flavours in the photon impact factor.  The parameters $A$ and $b$ are both given in units of GeV$^{-2}$.
\label{table:eval3}}
\end{table}
The final fit achieves  $\chi^2/N_{df} = 154.7/125 \sim 1.2$. This is a very good quality in view of the fact that the precision of the  data is very high, of the order of 2\%. The value of  $\chi^2/N_{df} \sim 1.2$ means that the precision of the theoretical computation is similar to data precision. This is remarkable in view of the fact that NLO corrections to the DIS impact factor  are missing. The KMS impact factor~\cite{KMS} which we are using takes into account the kinematical constraints, which are a part of the NLO correction, but not the complete correction.  The lack of full NLO corrections to the DIS impact factor could also be responsible for a slight worsening of the fit quality when the bottom flavour was added.  To support these arguments we also performed a fit in a higher $Q^2$ region ($Q^2>8$ GeV$^2$, $x<0.01$) and obtained a sizeable improvement in the quality of the five-flavour (four-flavour) fit, $\chi^2/N_{df} = 1.06$ ($1.00$).  The parameters of these fits are similar: $\kappa=0.63$ (0.64) and $b=17.1$ (18.2) GeV$^{-2}$ for the five-flavour (four-flavour) fit.

We also found that the fit is insensitive to the particular form of the proton impact factor as long as we allow the support to be concentrated close to $k_0$.  The proton impact factor does not explicitly alter the $x$-dependence as this is always encoded in the eigenvalues of the BFKL eigenfunctions.  However, the NLO corrections can introduce a scale such that the $x$-dependence associated with eigenfunction $n$ is amended from $x^{-\omega_n}$ to $(x/x_0)^{-\omega_n}$, thereby introducing another free parameter, $x_0$~\cite{Brodsky}.  We have examined this possibility and found that the best fits were nevertheless obtained with $x_0$ set to unity.

\begin{figure}[h]
\centerline{\epsfig{file=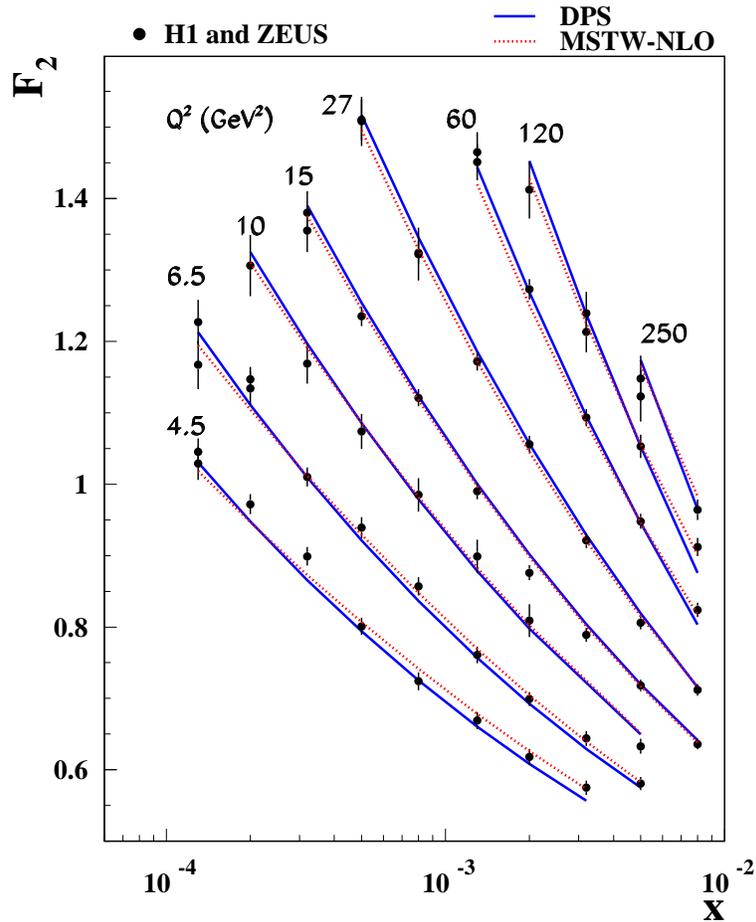, width=11 cm}}
\caption{The results of the fit, performed with 120 eigenfunctions and 30 overlap integrals for each eigenfunction (\emph{full line}), compared to a subsample of the low-$x$ HERA data~\cite{H1ZEUS}. The \emph{dotted line} shows the result of the NLO DGLAP global fit by MSTW~\cite{Thorne:2010kj}.}
\label{globfit} 
\end{figure}

The best DPS fit with five flavours (full line) is compared to a subsample of data~in~Fig.~\ref{globfit}.  In the same figure we also show the results of a standard NLO DGLAP global fit (dotted line) by MSTW~\cite{Thorne:2010kj}, which included the same combined HERA data~\cite{H1ZEUS}.  The MSTW fit~\cite{Thorne:2010kj} gives a somewhat better $\chi^2=112$ for the 128 low-$x$ HERA data points~\cite{H1ZEUS}, using around 30 free parameters\footnote{It is difficult to judge how many of these $\sim 30$ parameters in the DGLAP global fit are relevant to the low-$x$ HERA data, but a rough estimate is somewhere between five and ten.} in a global fit including many other data sets, so the quality of the fit to the low-$x$ HERA data is slightly better than the DPS fit.  Whereas in our formalism we see a considerable improvement in the quality of the fit if we restrict either $x<0.001$ or $Q^2>8$~GeV$^2$, in Ref.~\cite{Thorne:2010kj} the quality of the MSTW fit is similar over such restricted regions as it is for the whole region, $x < 0.01$ and $Q^2>4$ GeV$^2$.  The fits to the restricted regions give similar fit parameters as for the whole region and the dependence on the number of eigenfunctions, $n_{\rm max}$, shows a similar, slow convergence.  Therefore we consider the fit to the whole HERA low-$x$ region, for $Q^2>4$ GeV$^2$, as a representative fit and attribute a slight worsening of the quality of the fit to other effects which may become significant when $x$ is not sufficiently small, such as the effect of the neglected valence-quark contributions, or when $Q^2$ is not sufficiently large, such as NLO corrections to the photon impact factor (or possible effects from non-linear evolution).

\begin{figure}[h]
\centerline{\epsfig{file=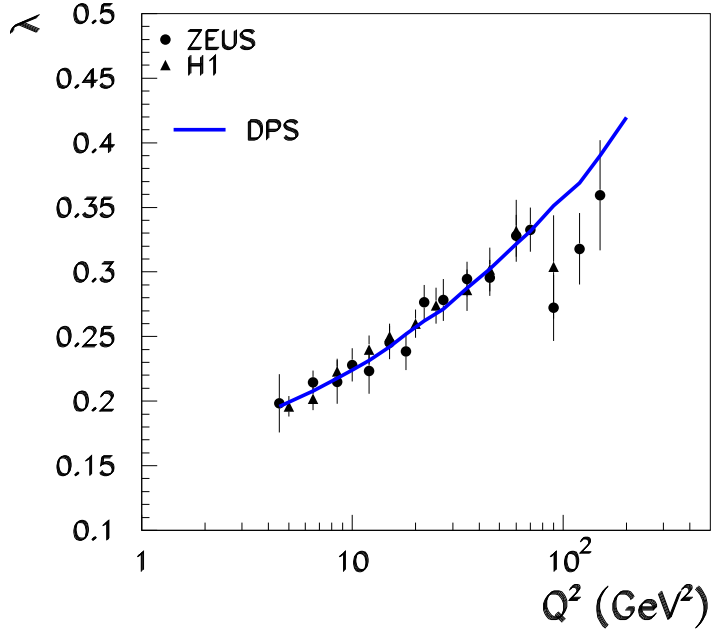, width=11 cm}}
\caption{The rate of rise $\lambda$, defined by $F_2\propto (1/x)^\lambda$ at fixed $Q^2$, as determined in the DPS fit and in the direct phenomenological fit to the data~\cite{H1ZEUSlam}.
} \label{fig-lam} 
\end{figure}

Figure~\ref{fig-lam} shows the rate of rise $\lambda$, defined by $F_2\propto (1/x)^\lambda$ at fixed $Q^2$, as determined in the DPS fit and in the direct phenomenological fit to the data~\cite{H1ZEUSlam}.  The present fit describes data very well.  This is a substantial improvement in comparison to our previous work~\cite{EKR}, where only a qualitative agreement with the observed $Q^2$ dependence of $\lambda$ was achieved.

\section{$\eta$--$\omega$ relation}

\begin{figure}[h]
\centerline{\epsfig{file=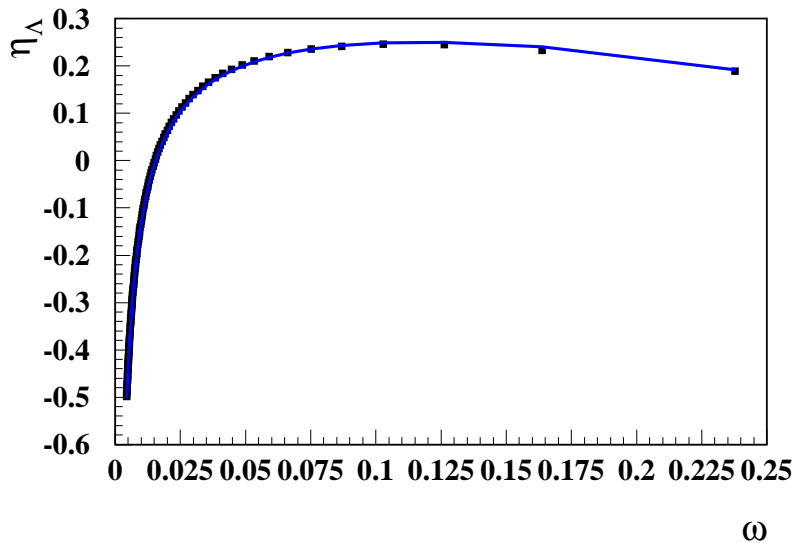, width=12 cm}}
\caption{Relation between the non-perturbative phase $\eta_\Lambda$ and the eigenvalues $\omega$.  The \emph{full line} shows the $\eta_\Lambda$--$\omega$ relation of eq.~(\ref{etarel}).
} \label{etaom} 
\end{figure}

The analogy with the Schr\"odinger equation discussed above suggests that perturbative wavefunctions can be smoothly extended to low $\log(k)$ values, i.e.~into the non-perturbative region. In this region an as-yet-unknown dynamics determines the values of the phase of wavefunctions which in turn determine the boundary conditions $\eta$.  The boundaries $\eta$ are determined at a low, but still perturbative, value of $k_0=0.3$ GeV\footnote{For future studies, it may be wiser to take a larger value of $k_0$ and use the infrared oscillation frequency to extrapolate the eigenfunctions to smaller values of momentum.  This extrapolation is a complicated process and so in this paper we have chosen this value of $k_0$ as a compromise value.} used in eq.~(\ref{boundary}).  The formalism described here allows a safe extrapolation of this phase to the non-perturbative region $k\approx \Lambda_{\rm QCD}$ by noting that the generalized phase $ \phi_\omega(k)$ can be expressed as~\cite{lipatov86}
\beq
 \phi_\omega(k)= 2\int^{k_{\rm crit}}_{k} \frac{d\,k'}{k'} |\nu_\omega(k')|
 \ = \ -2\nu_\omega(k) \ln(k)+2\int_0^{\nu_\omega(k)} \nu_\omega^{-1}(\nu^\prime) d\nu^\prime,   \label{phase3}
\eeq 
where in the integration by parts the relation $\nu_{\omega}(k_{\rm crit})=0$ was used.  In leading order, the integral over $d\nu^\prime$ is independent of $\omega$ and $k$.  In NLO this integral becomes $\omega$ dependent but is still independent of $k$, because it extends to the highest value of $\nu$, which does not depend on the value of $k_0$ (see Fig.~\ref{nuv-pap}).  This allows us to relate the phase $\eta$ defined at $k=k_0$ to the phase $\eta_\Lambda$ defined at $k=\Lambda_{\rm QCD}=0.220$~GeV by
\beq
\eta =-2\nu(k_0) \ln\left(\frac{k_0}{\Lambda_{\rm QCD}}\right) + \eta_\Lambda,
\label{boundary1}
\eeq
and to determine the $\eta_\Lambda$--$\omega$ relation shown in Fig.~\ref{etaom}.
We note that the exact value of $\Lambda_{\rm QCD}$ is somewhat arbitrary, depending on the order of perturbation taken for the $\beta$-function and the number of active flavours assumed.  However, any change in this value is absorbed by a small overall shift  in $\eta_\Lambda$. It does not affect our fit at all since in practice we choose a convenient value for $k_0$ and use $\eta$ rather than $\eta_\Lambda$.

It is instructive to express the $\eta$--$\omega$ dependence shown in Fig.~\ref{etaom} in an analytic form.  We have shown above that owing to the limited range of the frequency $\nu$, near $k=k_0$, we need to include a wide range of phases in order to be able to reproduce a proton impact factor which is highly suppressed for $k \, \gg \, \Lambda_{\rm QCD}$.  On the other hand we know that the eigenvalues, $\omega$, range from $\sim 0.25$ for the leading eigenfunction to zero, which is approached as we consider more and more eigenfunctions.  Since this is a small range of omega, it is inevitable that the $\eta$--$\omega$ relation will contain a singularity as $\omega \to 0 $.  Indeed we note that the approximate relation found in our analysis, $\omega_n = 0.5/(1+0.95\, n)$, when combined with the $\eta$--$n$ relation of eq.~(\ref{eta-n}), gives
\beq
\eta_\Lambda =0.4- \frac{0.0265}{\omega^{0.65}}, \;\;\;\;\;\;\;{\rm when} \;\;\;\; \omega \rightarrow 0.
\eeq
In addition to these asymptotic terms we found by fitting with a second order polynomial that
\beq
\eta_\Lambda =0.4 -0.14\omega - 1.9\omega^2 - \frac{0.0265}{\omega^{0.65}},
\label{etarel}
\eeq
which described the relation  between the non-perturbative phase $\eta_\Lambda$ and the eigenvalues $\omega$ very well, see Fig.~\ref{etaom}.

In leading-logarithmic approximation (LLA) the $ d\nu^\prime$ integral of eq.~(\ref{phase3}) can be evaluated analytically as
\beq 
\int_0^{\nu_0} \nu_\omega^{-1}(\nu^\prime) d\nu^\prime = \frac{4\pi}{\beta_0 \omega} \, a- \frac{\pi}{4} 
 \label{eq47} \eeq
 where $\beta_0=11-2n_f/3$ and $a=\int_0^{\nu_{0}} \chi_0(\nu^\prime)d\nu^\prime \approx 0.92$.  This led to the original assumption~\cite{lipatov86} of a constant phase, up to  $n\pi$ as in eq.~(\ref{boundary}), and gave the relation 
 \beq
 \omega_n= \frac{4a}{\beta_0} \,\, \frac{{1}} {n+\eta +1/4}.
\eeq

The fact that the RHS of eq.~(\ref{eq47}) diverges as $\omega \to 0$ is a reflection of the fact that as the eigenfunction number $n\to\infty$, $k_{\rm crit}\to \infty$, and the number of oscillations between $k_0$ and $k_{\rm crit}$ also becomes infinite.  The simple pole behaviour, however, is only valid in the approximation in which the characteristic function, $\chi$, is taken only to leading order and the coupling is run only in leading order. Once higher order terms are taken into account the constant behaviour of the non-perturbative phase $\eta$ is converted into a smooth behaviour and a power-like singularity (weaker than a perturbative pole).  The character of this singularity could be sensitive to the value of the $\beta$-function at any point between $k_0$ and $k_{\rm crit}$.  The singular term in eq.~(\ref{etarel}) (with a very small coefficient), obtained from our fit to data, is a reflection of the truncation of the perturbative expansion and the possible neglect of any BSM physics that may affect the $\beta$-function.

We recall that in the present evaluation the contributions of all known six quark flavours are taken into account but not the contribution from possible BSM particles such as superpartners.  The third eigenfunction, which has a critical momentum around 50 TeV, should already be sensitive to any such BSM physics.  This dependence could be substantial because the value of the $\beta$-function is changed significantly in the BSM region and more than 75\% of the phase integration, eq.~(\ref{phase}), extends over the BSM region, for $n>10$ eigenfunctions.    In other words, the phase at the lowest energy point $k_0$ is determined by integration over the whole  energy region, from $k_0$ up to $k_{\rm crit}$, and since the values of $k_{\rm crit}$ are very large (see Fig.~\ref{fig-kcrit}), the phase at $k_0$ is mainly determined by the energies beyond BSM thresholds.

The quantitative effects of BSM physics are very difficult to estimate because they change the $k$ dependence of the phases and the running of the coupling simultaneously. Therefore any detailed investigation of such effects requires a full NLO evaluation of the BSM effects both on the running of the coupling and the BFKL characteristic functions of the BFKL equation, which we plan to do in a forthcoming paper.

\section{Unintegrated gluon density}

\begin{figure}[h]
\centerline{\epsfig{file=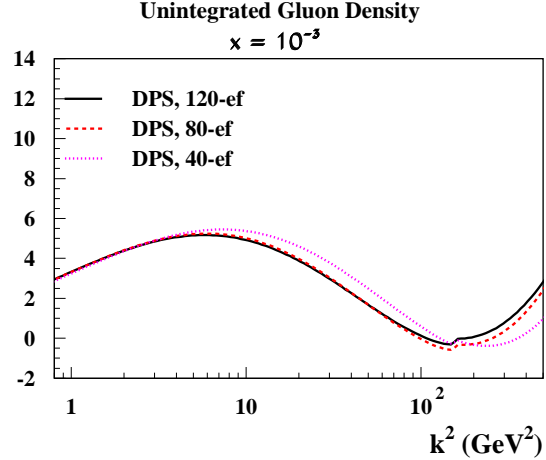, width=9 cm}}
\centerline{\epsfig{file=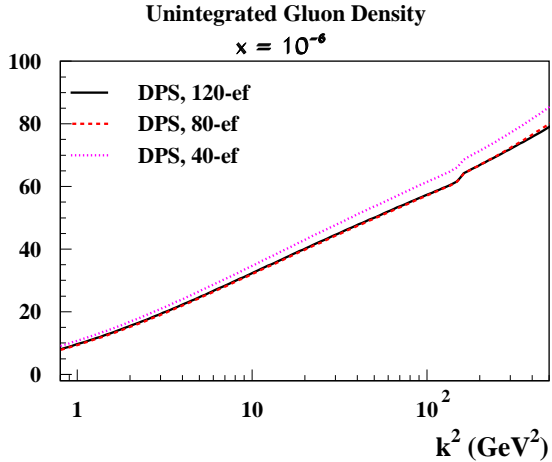, width=9 cm}}
\caption{Unintegrated gluon density at $x=10^{-3}$ and $x=10^{-6}$.} \label{un-glu2} 
\end{figure}

The unintegrated gluon density, $xg(x,k)$, is defined in analogy to eq.~(\ref{un-glu}) by
\beq  xg(x,k) \ = \   k^2
\sum_{m,n}   \left( \frac{k}{x} \right)^{\omega_n}  f_n(\mathbf{k})  {\cal N}^{-1}_{nm}  {\cal A}^{(D)}_m. \eeq
Figure~\ref{un-glu2} shows the unintegrated gluon density at two values of $x$,  $x=10^{-3}$ and $x=10^{-6}$. The unintegrated gluon density determined in the discrete BFKL approach is positive for all values of $k$ in the HERA data region. In addition to the  results based on fits with 120 eigenfunctions the figure shows also the results of fits with 40 and 80 eigenfunctions. At  $x=10^{-3}$, the fits with 80 (40) eigenfunctions start to differ sizeably from the 120 eigenfunction fit at $k^2\sim 100 \;  (10)$ GeV$^2$. The fit with 80 (40) eigenfunctions has $\chi^2 = 156 \; (193)$ for $N_{df}=125$ (see Table~\ref{table:eval1}), which means that the differences from the best fit are very small for 80 eigenfunctions and even for 40 eigenfunctions these differences can be hardly seen on the scale of Fig.~\ref{globfit}.  The fact that differences between the fits with the lower number of eigenfunctions can easily be seen in the unintegrated gluon density, at  $x=10^{-3}$,  but not in $F_2$ means that the contribution of higher $n$ eigenfunctions are more important in the unintegrated gluon density than in $F_2$.  With the deceasing $x$  the convergence  with  increasing number of eigenfunctions  improves. At  $x=10^{-6}$, the unintegrated gluon density of the fit with 80 eigenfunctions is almost indistinguishable from  the 120 eigenfunction fit. The differences between the 40 and 120  eigenfunction fit, although smaller than at  $x=10^{-3}$, are still visible, however.  Therefore, HERA data analysed in the discrete BFKL schema are determining the unintegrated gluon density up to $k^2< {\cal O}(100)$ GeV$^2$.  For higher $k^2$ one should use either more eigenfunctions or change to DGLAP evolution for $k^2>{\cal O}(100)$ GeV$^2$.

The very good description of the HERA $F_2$ data in the discrete pomeron scheme requires a large number of eigenfunctions.  This has several reasons.  First we recall that the enhancement factor of the leading pole, $x^{-\omega_1}$, is not very large, e.g.~at $x=10^{-3}$ it is only around a factor of five, because $\omega_1\approx 0.25$.  Then the contributions of large $n$ eigenfunctions  are only weakly suppressed in the Green function; the only suppression factor is due to  normalization which requires  integration over the $k$ region proportional to $\log(k_{\rm crit})$. This region grows almost linearly with  $n$, which leads to a suppression of the amplitude of the oscillations of eigenfunction number $n$ as $\sim 1/\sqrt{n}$.  This, in turn, leads to a slow decrease (also $\sim1/\sqrt{n}$) of the coefficients of these eigenfunctions required for the fit.  Since the proton impact factor has a narrow range of support in $k$ it has a large bandwidth and therefore requires a large number of eigenfunctions in order to describe with sufficient accuracy the high quality HERA data.

The large number of eigenfunctions required to obtain a good fit to HERA data could indicate that the basic QCD interactions which determine the BFKL kernel should be modified. Indeed the values of $k_{\rm crit}$ cross the postulated supersymmetry thresholds ($\sim10$~TeV) between $n=2$ and $n=3$ and they approach the Planck scale at $n\approx 10$.  This strongly suggests that the characteristic function at these high energies should be substantially changed by crossing  thresholds for new physics.  The effects of such changes are difficult to estimate without performing an evaluation in which the BFKL kernel is modified by N=1 to N=4 supersymmetry effects~\cite{kotlip}.

At first sight, the suggestion that there can be signals for BSM physics from deep-inelastic scattering at energies significantly below the threshold for such new physics appears to be in conflict with the general principle of decoupling.  However, it is perfectly possible to gain information about thresholds for new physics from ``low-energy'' data coupled with an assumption about some high-energy behaviour, generated by the running of couplings. An analogy is the prediction of new physics thresholds from the LEP measurements of gauge-couplings and the postulate of coupling unification at some GUT scale---these being only compatible if the running of the couplings are amended by new physics.  In our case, it can be argued that the application of the BFKL formalism, which implies a critical momentum for which the oscillation frequency vanishes, and the fit at ``low-energy'', can yield information about the running of the QCD coupling between these scales, such that the presence of new physics thresholds affects the quality of the fit.  It is the infrared phases (universal to all processes mediated by a pomeron)\footnote{At present the only precise data at small enough values of $x$ for the BFKL formalism to be applicable are from HERA structure functions.  Alternative processes such as low-mass forward Drell--Yan production, to be studied at the LHC, may also provide data at the necessary low values of $x$.} that would be affected by such BSM physics and, as explained above, such a sensitivity is permitted within the context of the decoupling theorem as it is generated by the running of the coupling over a large range. The proton impact factor would {\it not} display such sensitivity.  Since this is unknown, it is also possible that the effect of the change in the phases of the BFKL eigenfunctions would be partially compensated by a change in the parameters that we use to describe the proton impact factor.  Nevertheless, it is perfectly plausible that the quality of the overall fit could be sensitive to BSM physics and therefore provide some signal, since the shape of the proton impact factor is fairly tightly constrained by the requirement of positivity and narrow support.  Any detailed investigation of these effects requires a full NLO evaluation of BSM effects which we plan to do in a forthcoming publication.

\section{Comparison with gluon distribution from a DGLAP analysis}

Great care must be exercised if one wishes to attempt to compare the gluon distribution obtained here with that obtained from a DGLAP analysis~\cite{MSTW} or even a DGLAP analysis supplemented with BFKL dynamics at low $x$~\cite{TW}.
\begin{enumerate}
\item In the first place, the comparison can only be made under the assumption that one is working at sufficiently low $x$ for the structure functions to be dominated by the gluon distribution alone.  Although this will eventually be the case if $x$ is sufficiently small, in the HERA region the quark distribution used in the DGLAP analyses remains numerically important. The gluon contribution is suppressed by an overall factor of $\alpha_s(Q^2)$ and also (substantially) by the convolution of the gluon coefficient function $C_g(x)$ with the gluon density.  The separation of the quark and gluon distributions depends on the initial values assumed in the DGLAP analysis at some reference value of $Q^2$. In the BFKL approach, it is assumed that there is {\it no} primordial quark density emerging from the proton and that quarks appear (in NLO BFKL) only from pair production from a gluon somewhere along the ladder~\cite{FL}. This is substantially different from the fitted densities extracted from the DGLAP analysis from data used to obtain the distributions in~\cite{MSTW,TW}.
\item In the BFKL analysis, the counterpart of the gluon coefficient function is the upper impact factor $\Phi_{\mathrm{DIS}}(x,k,Q)$.\footnote{In leading order this impact factor is $x$-independent, but we use the more careful analysis of~\cite{KMS} in which the discrepancy between the measured Bjorken-$x$ and the longitudinal momentum of the unintegrated gluon density is accounted for.}  The equivalence between this formalism and the DGLAP approach is obtained within the approximation
\beq \int \frac{dk}{k} g(x,k^2) \otimes \Phi_ {\mathrm{DIS}}(x,k,Q) \ \approx \ 
 G(x,Q^2) \otimes C_g(x) + \alpha_s(Q^2) P_{qg}(x) \otimes G(x,Q^2) \ln\left(\frac{Q^2}{k^2}\right), \eeq
where $G(x,Q^2)$ is the integrated gluon density, $P_{qg}$ is an off-diagonal Altarelli--Parisi splitting function, and $\otimes$ indicates a convolution in $x$.

This approximation is obtained in the leading-logarithm approximation, i.e.~for sufficiently large $Q^2$ and the equivalence can be demonstrated in the continuum BFKL case (see e.g.~\cite{FR}) in the double-leading-logarithm limit, in which the integral over all frequencies, $\nu$, is dominated by the saddle point in the complex $\nu$ plane. In our analysis, we may indeed simulate the continuum case for the higher eigenfunctions, whose eigenvalues are very closely spaced. However, in order to ensure that the above-mentioned saddle point is captured in our discrete sum, we would need to take many more discrete eigenfunctions, that number increasing with increasing $Q^2$. This is demonstrated by the fact that as we go to higher values of $Q^2$ we need a larger number of eigenfunctions in order to obtain an integrated gluon density which is stable (convergent) and everywhere positive.

\item Finally, there is the matter of the renormalization prescription dependence of the anomalous dimensions themselves, i.e.~between the $\overline{\rm MS}$ usually used in a DGLAP analysis and the BFKL prescription to which we are forced in a direct comparison of the BFKL analysis with data.  This was first considered in~\cite{CH} and developed in~\cite{CC}, where a translation of the two prescriptions was given which is valid for transverse momenta above $k_{\rm crit}$ but it is unclear how to continue this into the oscillatory region.  In the analysis of~\cite{TW} this matter was left as an open question.    
\end{enumerate}
We therefore content ourselves with the fact that we have obtained a gluon distribution, relevant to our pure BFKL approach, valid up to $Q^2=100 \, \mathrm{GeV^2}$, which fits HERA data very well but which does {\it not} lend itself to a comparison with the distributions obtained from a DGLAP analysis.

\section{Summary and conclusions}

In this paper we have shown that NLLA solutions of the BFKL equation with the running coupling describe all properties of HERA $F_2$ data very well, for $Q^2>4$ GeV$^2$ and $x<10^{-2}$, provided we allow the infrared phase, $\eta$, to vary with the eigenvalues $\omega_n$.  The solutions of this equation have oscillatory form, in which the  frequencies $\nu(k^2)$ are varying due to the running of $\alpha_s(k^2)$.  We solve the equation near the point $\nu=0$ which singles out a specific value of $k=k_{\rm crit}$, where $\nu(k_{\rm crit}^2)=0$.  We show that the solutions of the BFKL equation obtained here can be considered as a quantized version of the solutions of the DGLAP equation.  The matching of the BFKL solutions in the region $k \, \sim \, k_{\rm crit}$ leads to a unique set of discrete eigenfunctions which cannot be obtained from the  DGLAP equation alone.
 
The description of data is obtained by convoluting the Green function with the photon and the proton impact factors.  The photon impact factor is known, while the shape of the proton impact factor was assumed to follow a simple exponential form.  The comparison with data shows that a particular functional form of the proton impact factor is not very important as long as it is positive and concentrated at the values of $k < {\cal O}(1)$ GeV.  The limited support of the proton impact factor requires, however, a convolution with a large number of the eigenfunctions, subjected to a specific phase condition which was determined from the fit to data.
 
The fitting procedure, especially the finding of the proper infrared phase condition, was only possible because of recently published combined H1 and ZEUS $F_2$ data from HERA.  The increased precision of this data requires a large number of eigenfunctions, up to 120, to obtain a good fit.  The resulting fit permits a very good description of the $F_2$ data and the $Q^2$ dependence of the exponent, $\lambda$, of $1/x$, for small-$x$.  The resulting gluon density is positive, in the range of HERA energies.

At higher energies the resulting gluon density is sensitive to the number of eigenfunctions used in the fit.  Since the higher eigenfunctions have eigenvalues which become closer together, the inclusion of such eigenfunctions effectively simulates a continuum on top of the first few discrete (clearly separated) ones. This could indicate that  we are approaching a continuum limit which could be better described by the DGLAP evolution alone. So the BFKL solution could determine the gluon density up to $k^2$ of the order of ${\cal O}(100)$ GeV$^2$, and from then on the DGLAP solution could be used. This could provide a method to overcome the problem of negative gluon densities at low $x$ and small scales pertinent to the standard DGLAP fits.  For example, in Ref.~\cite{TW}, in contrast to the standard DGLAP result, the input gluon at $Q_0^2 = 1$~GeV$^2$ obtained from a global fit including small-$x$ resummation was positive and slightly increasing as $x\to0$.  In view of the importance of the gluon density to the LHC physics we plan to study this issue in a forthcoming paper.

The solutions of the BFKL equation together with HERA data determine the relation between the eigenvalue $\omega$ and the phase $\eta_\Lambda$ which consists of a polynomial term and a singular term in $\omega$.  The polynomial term contains information about the non-perturbative gluonic dynamics inside the pomeron because we show that the BFKL  equation can be considered to be analogous to the Schr\"odinger equation for the wavefunction of the (interacting) two-gluon system.  The BFKL kernel corresponds to the Hamiltonian with the eigenvalues $\omega_n$.  The analogy with the Schr\"odinger equation  suggests that  perturbative wavefunctions can be  smoothly extended to very low virtuality values, $k^2$, i.e.~into the non-perturbative region. In this region an as-yet-unknown dynamics determines the values of the phase of wavefunctions which in turn determine the boundary conditions $\eta_\Lambda$.

The singular term, on the other hand, is presumably generated by the perturbative effects which were not fully taken into account in our evaluation.  This term is sensitive to the high virtuality behaviour of the gluon--gluon amplitude, much beyond the virtualities which are actually tested in the experiment.  This remarkable property is due to the fact that in the evolution scheme developed here, the BFKL equation is solved near the critical point, $k_{\rm crit}$, and the value of $k_{\rm crit}$ grows quickly with the increase of the eigenfunction number, rapidly crossing proposed thresholds for new physics and even the Planck scale.  Since we found that the proper description of data requires a large number of eigenfunctions we obtain an apparent sensitivity to the BSM effects.  We recall that in our approach the eigenfunctions with large $n$ correspond to Regge poles which are similar to hadrons with a very small size, because $k_{\rm crit}$ is very large.  Their masses (i.e.~$\omega _n$) are small but can depend on the BSM physics.  The sheer potential existence of BSM particles, although never produced in the interactions relevant to the fitted data, modify the running of the coupling and the (NLO) characteristic function of the BFKL equation below the critical momenta and, in turn, modify the frequency, amplitude and phase of the eigenfunctions at low-energy.  We have shown that these states have a soft hadronic tail (i.e.~a part of the eigenfunctions around $k\sim\Lambda_{\rm QCD}$) by which they interact with proton and photon and give an essential contribution to the structure functions.

However, only a full NLLA evaluation which takes into account all possible BSM states can show whether this apparent sensitivity turns out to be real.  If it turns out to be sufficient to act as a signal for BSM physics this would provide a new method of ``telescoping the Planck scale''~\cite{zerwas}.


\begin{thebibliography}{99}


\bibitem{H1ZEUS}

  F.~D.~Aaron {\it et al.}  [H1 and ZEUS Collaborations],
  JHEP {\bf 1001} (2010) 109.


\bibitem{BFKL} 

  I.~I.~Balitsky and L.~N.~Lipatov,
  Sov.\ J.\ Nucl.\ Phys.\  {\bf 28} (1978) 822;

  E.~A.~Kuraev, L.~N.~Lipatov and V.~S.~Fadin,
  Sov.\ Phys.\ JETP {\bf 44} (1976) 443;

 V.~S.~Fadin, E.~A.~Kuraev and L.~N.~Lipatov,
 Phys.\ Lett.\  B {\bf 60} (1975) 50.


\bibitem{FL}

  V.~S.~Fadin and L.~N.~Lipatov,
  Phys.\ Lett.\  B {\bf 429} (1998) 127;

  M.~Ciafaloni and G.~Camici,
  Phys.\ Lett.\  B {\bf 430} (1998) 349.


\bibitem{salam}

  G.~P.~Salam,
  JHEP {\bf 9807} (1998) 019.


\bibitem{EKR}

  J.~Ellis, H.~Kowalski and D.~A.~Ross,
  Phys.\ Lett.\  B {\bf 668} (2008) 51.


\bibitem{lipatov86}
  L.~N.~Lipatov,
  Sov.\ Phys.\ JETP {\bf 63} (1986) 904.


\bibitem{GLR}

  L.~V.~Gribov, E.~M.~Levin and M.~G.~Ryskin,
  Phys.\ Rept.\  {\bf 100} (1983) 1.


\bibitem{levin98}


  E.~Levin,
  Nucl.\ Phys.\  B {\bf 545} (1999) 481.


\bibitem{DGLAP}

  V.~N.~Gribov and L.~N.~Lipatov,
  Sov.\ J.\ Nucl.\ Phys.\  {\bf 15} (1972) 438;

  L.~N.~Lipatov,
  Sov.\ J.\ Nucl.\ Phys.\  {\bf 20} (1975) 94;

  G.~Altarelli and G.~Parisi,
  Nucl.\ Phys.\  B {\bf 126} (1977) 298;

  Y.~L.~Dokshitzer,
  Sov.\ Phys.\ JETP {\bf 46} (1977) 641.


\bibitem{KMS}

  J.~Kwiecinski, A.~D.~Martin and A.~M.~Stasto,
  Phys.\ Rev.\  D {\bf 56} (1997) 3991.


\bibitem{trimvirate}

  Y.~V.~Kovchegov and H.~Weigert,
  Nucl.\ Phys.\  A {\bf 784} (2007) 188;

  E.~Levin,
  Nucl.\ Phys.\  B {\bf 453} (1995) 303.


\bibitem{Brodsky}

  S.~J.~Brodsky, V.~S.~Fadin, V.~T.~Kim, L.~N.~Lipatov and G.~B.~Pivovarov,
  JETP Lett.\  {\bf 76} (2002) 249.


\bibitem{Thorne:2010kj}
  R.~S.~Thorne, A.~D.~Martin, W.~J.~Stirling and G.~Watt,
  arXiv:1006.2753 [hep-ph].


\bibitem{H1ZEUSlam}

  C.~Adloff {\it et al.}  [H1 Collaboration],
  Eur.\ Phys.\ J.\  C {\bf 21} (2001) 33;

  S.~Chekanov {\it et al.}  [ZEUS Collaboration],
  Eur.\ Phys.\ J.\  C {\bf 21} (2001) 443.


\bibitem{kotlip}

  A.~V.~Kotikov and L.~N.~Lipatov,
  Nucl.\ Phys.\  B {\bf 582} (2000) 19;

  Nucl.\ Phys.\  B {\bf 661} (2003) 19.


\bibitem{MSTW}

  A.~D.~Martin, W.~J.~Stirling, R.~S.~Thorne and G.~Watt,
  Eur.\ Phys.\ J.\  C {\bf 63} (2009) 189.


\bibitem{TW}

  C.~D.~White and R.~S.~Thorne,
  Phys.\ Rev.\  D {\bf 75} (2007) 034005.


\bibitem{FR}

  J.~R.~Forshaw and D.~A.~Ross,
  \emph{Quantum Chromodynamics and the Pomeron},
  Cambridge Lect.\ Notes Phys.\  {\bf 9} (1997) 1.


\bibitem{CH}

  S.~Catani and F.~Hautmann,
  Nucl.\ Phys.\  B {\bf 427} (1994) 475.


\bibitem{CC}

  M.~Ciafaloni, D.~Colferai, G.~P.~Salam and A.~M.~Stasto,
  Phys.\ Lett.\  B {\bf 635} (2006) 320;

  S.~Catani, M.~Ciafaloni and F.~Hautmann,
  Phys.\ Lett.\  B {\bf 307} (1993) 147.


\bibitem{zerwas}
  P.~Zerwas, ``High-energy physics: Telescoping the Planck scale'', Farewell Colloquium for Rolf-Dieter Heuer, 5 December 2008, \texttt{http://heuer-colloquium.desy.de/e9/}.


\end{thebibliography}
\end{document}